\documentclass[rnote]{aa}

\usepackage{graphicx}
\usepackage{txfonts}

\begin{document}

   \title{On the origin and evolutionary state of RZ~Cas, KO~Aql and S~Equ}

   \author{N. Mennekens\inst{1}
          \and
          J.-P. De Greve\inst{1}
          \and
          W. Van Rensbergen\inst{1}
          \and
          L. R. Yungelson\inst{2}
          }

   \institute{Astrophysical Institute, Vrije Universiteit Brussel, Pleinlaan 2, 1050 Brussels, Belgium\\
             \email{nmenneke@vub.ac.be, jpdgreve@vub.ac.be, wvanrens@vub.ac.be}
             \and
             Institute of Astronomy of the Russian Academy of Sciences, 48 Pyatnitskaya St., 119017 Moscow, Russia\\
             \email{lry@inasan.ru}
             }

   \date{Received February 7, 2008; accepted April 30, 2008}

   \abstract
   {}
   {Determination of the present evolutionary state and the restrictions on the initial mass ratios of RZ~Cas, KO~Aql and S~Equ.}
   {Comparison of mass gaining stars with evolutionary models of single stars with the same mass and subsequent comparison with accretion tracks from simultaneous conservative binary evolution.}
   {The gainers are in an early main sequence stage ($X_c > 0.5$), with KO~Aql being almost unevolved (assuming quasi-thermal equilibrium). The initial donor/gainer mass ratios $M_{di} / M_{gi}$ must be larger than three to obtain the present mass and luminosity of the gainers.}
   {}

   \keywords{stars: binaries: eclipsing --
                stars: binaries: close --
                stars: individual: RZ~Cas, KO~Aql, S~Equ
               }

   \maketitle

\section{Introduction}

\begin{table*}
\caption{Orbital and physical parameters of RZ~Cas, KO~Aql and S~Equ}
\label{table:parameters}
\centering
\begin{tabular}{c c c c c c c}
\hline\hline
Parameter & \multicolumn{2}{c}{RZ~Cas} & \multicolumn{2}{c}{KO~Aql} & \multicolumn{2}{c}{S~Equ} \\
 & gainer & donor & gainer & donor & gainer & donor \\
\hline
$P \ (d)$ & \multicolumn{2}{c}{1.195} & \multicolumn{2}{c}{2.864} & \multicolumn{2}{c}{3.436} \\
$M \ (M_{\odot})$ & $2.21 \pm 0.08$ & $0.73 \pm 0.02$ & $2.53 \pm 0.05$ & $0.55 \pm 0.01$ & $3.24 \pm 0.03$ & $0.42 \pm 0.01$ \\
$R \ (R_{\odot})$ & $1.67 \pm 0.03$ & $1.94 \pm 0.03$ & $1.74 \pm 0.07$ & $3.34 \pm 0.07$ & $2.74 \pm 0.09$ & $3.24 \pm 0.10$ \\
$\mathrm{log} \ T \ (K)$ & 3.934 & 3.672 & 3.996 & 3.646 & 4.049 & 3.721 \\
$\mathrm{log} \ L/L_{\odot}$ & $1.12 \pm 0.02$ & $0.16 \pm 0.08$ & $1.41 \pm 0.06$ & $0.56 \pm 0.06$ & $2.02 \pm 0.10$ & $0.86 \pm 0.10$ \\
$\dot{M} \ (M_{\odot}/yr)$ & \multicolumn{2}{c}{$\sim 1.3 \cdot 10^{-6}$} & \multicolumn{2}{c}{$\sim 2.6 \cdot 10^{-7}$} & \multicolumn{2}{c}{$\sim 4.0 \cdot 10^{-8}$} \\ 
\hline
\end{tabular}
\end{table*}

RZ~Cas, KO~Aql and S~Equ are semi-detached eclipsing Algol-type binaries for which we investigate restrictions on their origin coming from the comparison of observations with theoretical evolution models. For RZ~Cas we use the orbital and physical parameters compiled by \cite{maxted94}, and the $\dot{M}$ given by \cite{lehmann07}. The $\dot{M}$ for RZ~Cas was derived from O-C curves. For KO~Aql and S~Equ, we use the results of \cite{soydugan07}. All these are rendered in Table~1. To avoid any confusion, we label the hotter, mass-gaining component (commonly called the primary) as \emph{gainer}, while the cooler, mass-losing component (the secondary) is labeled as \emph{donor}. The mentioned authors conclude from their analysis that the three systems are showing prominent features of mass transfer. Together they form an interesting set from an evolutionary point of view. The total masses range from $2.94 M_{\odot}$, over $3.08 M_{\odot}$ to $3.66 M_{\odot}$, the mass ratios from 0.13 over 0.22 to 0.33. The total mass, gainer mass and period of RZ~Cas are the smallest, but combine with the largest mass ratio, whereas this is the opposite for S~Equ. KO~Aql lies in the middle of the two. This might indicate that the three systems are in subsequent stages of mass transfer. The low mass ratios point towards an advanced state of mass transfer. 

\section{Comparison with single star evolution}

A comparison of the location of the gainers with single star evolutionary tracks can teach us how far they deviate from normal stars. Single star evolution (initial composition: $X = 0.7$, $Z = 0.02$) was computed for the three masses of the gainers in Table~1 with the same evolutionary code used by \cite{deloore05}. The results are shown in Figure~1, together with the positions of the three mass gaining stars. The mass of the lowest track corresponds to that of RZ~Cas, the next to that of KO~Aql, and the highest to the mass of S~Equ.

It is clear that the observed gainers are underluminous compared to the theoretical counterparts, as was previously noted by \cite{soydugan07}. The evolution of a close binary can be divided into two distinct phases: an initial phase of fast mass transfer (increasing $\dot{M}$) on a thermal timescale, and a later phase of slower mass transfer (decreasing $\dot{M}$) on a nuclear timescale (e.g. advanced state with well reversed mass ratio).

\begin{figure*}
   \centering
   \includegraphics[width=488px]{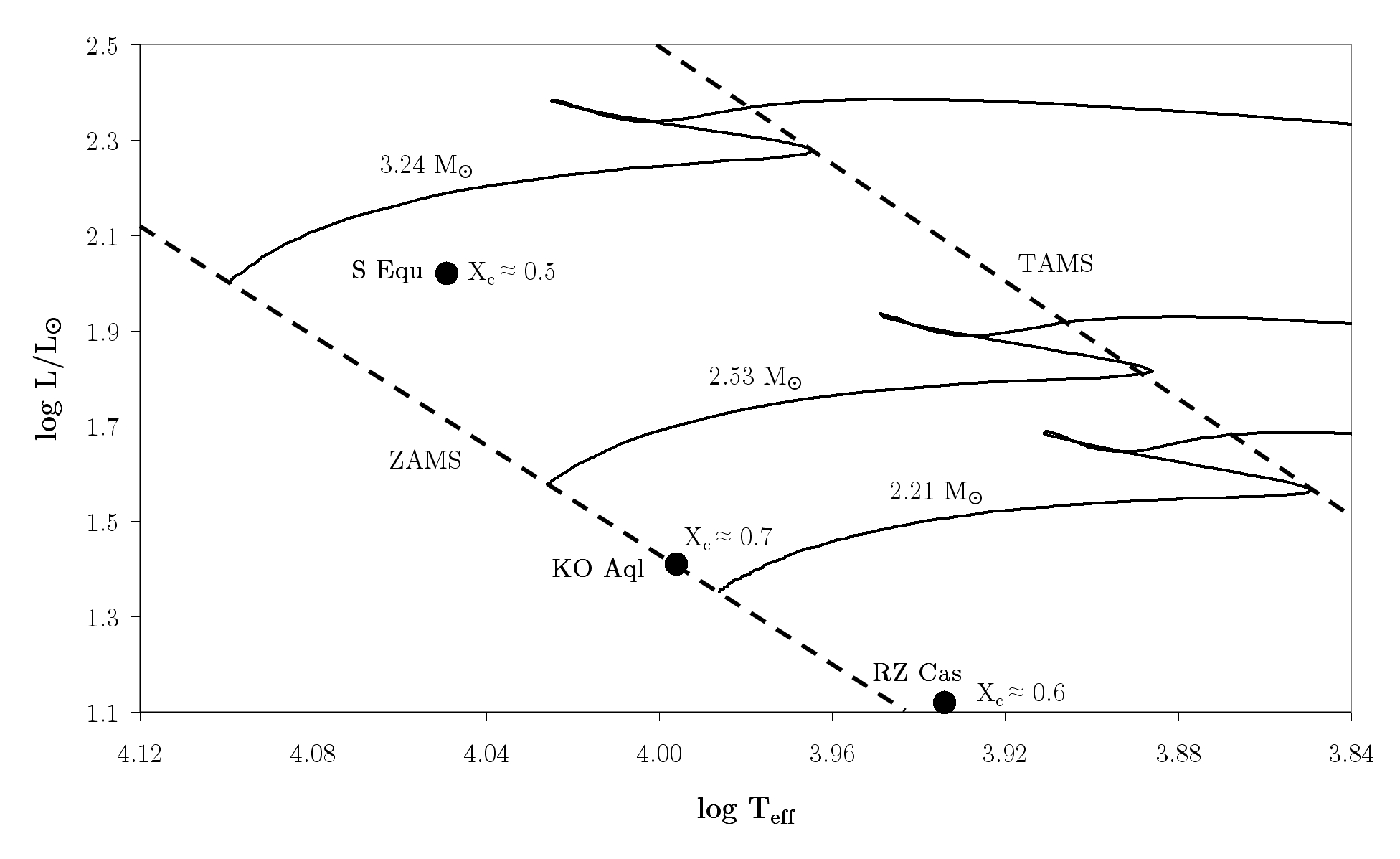}
   \caption{HRD with the positions of the gainers of S~Equ, KO~Aql, RZ~Cas, and tracks of single star evolution for the mass values of Table~1. Central hydrogen abundance by mass, as interpolated from the models, is indicated next to the gainers (see text).}
   \label{Fig1}
\end{figure*}

Given the well reversed mass ratios (RZ~Cas: 0.33; KO~Aql: 0.22; S~Equ: 0.13) and the relatively low mass loss rates deduced from the period changes ($\sim 10^{-6}$ to $10^{-8} M_{\odot}/yr$), we assume that the systems are well beyond the peak phase of mass transfer. In that case, the gainer is close to thermal equilibrium and the evolution is mostly determined by the inner structure. Thus, its position in the main sequence band reflects the amount of nuclear burning that has taken place and we can derive a rough estimate for the amount of hydrogen by mass in the core ($X_c$). We find that for RZ~Cas $X_c \approx 0.6$, for KO~Aql $X_c \approx 0.7$, and for S~Equ $X_c \approx 0.5$. In conclusion, this points towards a ZAMS structure for KO~Aql, and a more evolved structure for S~Equ.

\section{Conservative considerations}

To explore the values for the masses and the period of the progenitor systems, we first investigate a possible conservative history of mass transfer (conservation of mass and angular momentum). We specifically discuss the accretion behavior in some detail for the middle system KO~Aql (in mass, mass ratio and period), and we will extend the conclusions to the other two. In the code, the structure of the gainer is calculated assuming that the added matter has the same entropy as the gainer and neglecting all effects of interaction.

We calculated the system parameters of KO~Aql backwards to initial states corresponding to $q_i = 1.01$ (initial gainer mass $M_{gi} = 1.53 M_{\odot}$), 1.32 ($M_{gi} = 1.32 M_{\odot}$), and 1.61 ($M_{gi} = 1.18 M_{\odot}$), subsequently followed by a computation of the evolution of the gainers (through simultaneous evolution of both components). All three failed to reproduce the current parameters of the gainer of KO~Aql, in mass as well as in HRD position. The calculations were stopped when the mass transfer rate fell below $10^{-11} M_{\odot}/yr$, and nuclear evolution took over (and the donor returned to thermal equilibrium). The mass of the gainer at that moment for all three systems was in the range $1.80 M_{\odot}$ to $1.86 M_{\odot}$. The results for the gainer in the HRD are shown in Figure~2.

\begin{figure*}
   \centering
   \includegraphics[width=488px]{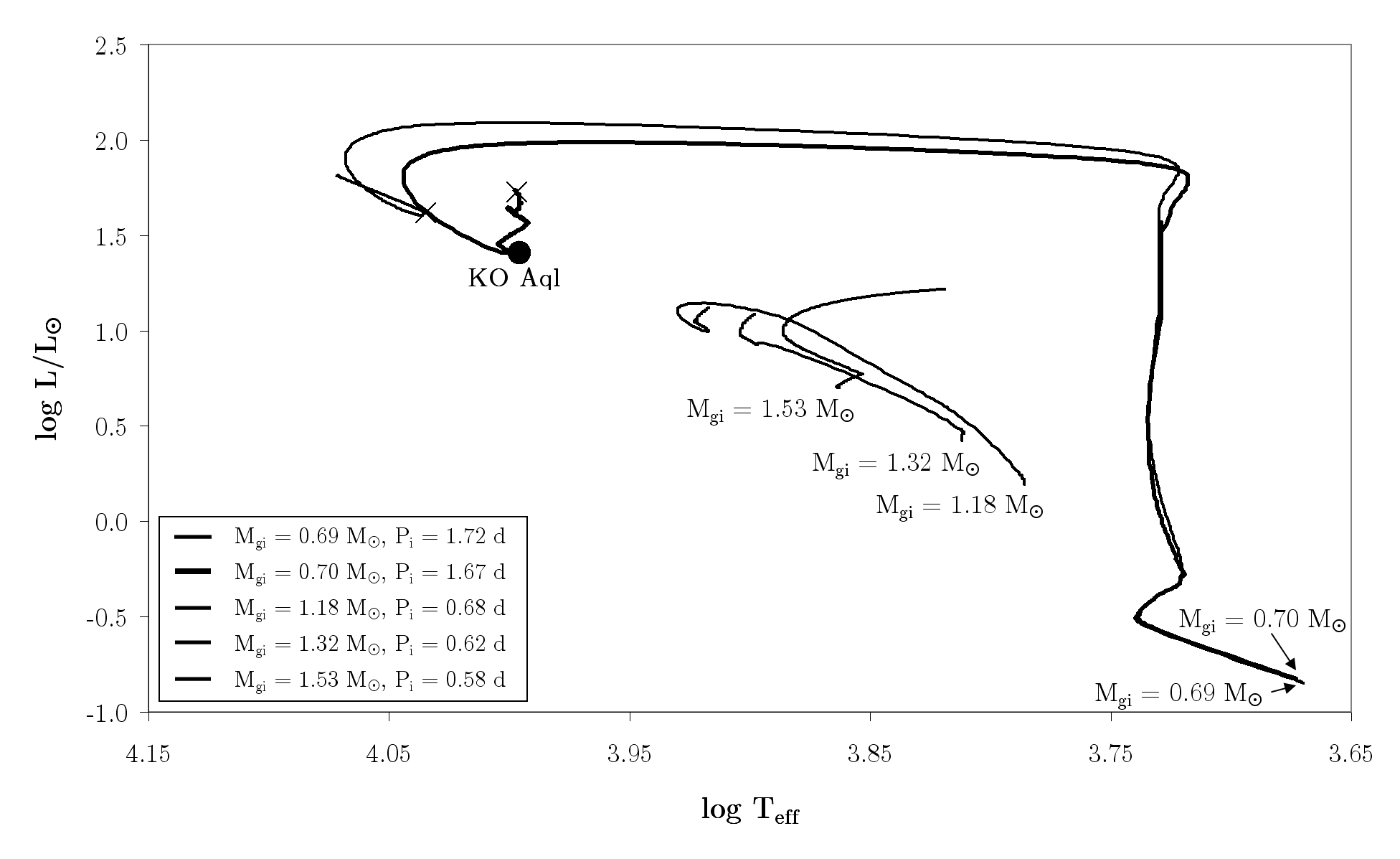}
   \caption{HRD with accretion tracks for the gainer of KO~Aql. The initial mass is indicated at the beginning of the track. Crosses indicate the positions where the masses and period of the systems equal those of KO~Aql.}
   \label{Fig2}
\end{figure*}

All three resulted in gainers with masses and HRD positions far below the observed one. We note that the larger the initial mass ratio, the larger is the mass of the gainer and its luminosity in the slow phase. Therefore, we calculated a new set of two with more extreme mass ratios (3.40, $M_{gi} = 0.70 M_{\odot}$ and 3.46, $M_{gi} = 0.69 M_{\odot}$). The results are also shown in Figure~2. Both tracks show a similar behavior. As a result of the large difference in thermal equilibrium timescales between the components ($t_{KH,g}/t_{KH,d} \sim 3 - 5$), the star first strongly deviates from thermal equilibrium, and strongly increases in volume. After mass ratio reversal the mass transfer rate decreases and the gainer evolves at nearly constant luminosity towards the ZAMS. Near the ZAMS the star resumes quasi-thermal equilibrium and moves downward along the ZAMS, finally settling itself in the slow phase of mass transfer on a track corresponding to its mass and inner structure. The mass transfer rate evolves from a fast phase of $\sim 1.7 \cdot 10^{-6} yr$ with a maximum of a few times $10^{-6} M_{\odot}/yr$ to an average of $10^{-10} M_{\odot}/yr$ in the rest of the main sequence phase. The first series ($M_{gi} = 0.70 M_{\odot}$) results in an accretion track that runs through the observed position of KO~Aql in the HRD. At that position it does not result in a mass $M_g$ matching the observed one. Indeed, at the moment that it reaches the HRD position of KO~Aql, its mass is $2.25 M_{\odot}$. During the further evolution, the mass transfer (of about $\dot{M} \sim 10^{-7} - 10^{-8} M_{\odot}/yr$) increases that mass to $\sim 2.54 M_{\odot}$. At $M = 1.53 M_{\odot}$, it is already well evolved into the main sequence, its location corresponding to single star models with the same $X_c$.

The second track ($M_{gi} = 0.69 M_{\odot}$) lies well above the observed position, but the higher mass loss rate results in the mass of KO~Aql and even beyond. The characteristics of the model at the moment that $M_g = 2.53 M_{\odot}$ are $\mathrm{log} \ T_{eff} = 4.035$ and $\mathrm{log} \ L / L_{\odot} = 1.618$. For the two calculations that attain the masses and period of KO~Aql, the calculated properties at that time are given in Table~2.

\begin{table*}
\caption{Parameters of the different calculations when they have the masses and period of KO~Aql.}
\label{table:KOAql}
\centering
\begin{tabular}{c c c}
\hline\hline
$M_{gi}$ & $0.69 M_{\odot}$ & $0.70 M_{\odot}$ \\
\hline
$\mathrm{log} \ T_g \ (K)$ & 4.035 & 3.997 \\
$\mathrm{log} \ L_g/L_{\odot}$ & 1.618 & 1.727 \\
$\dot{M} \ (M_{\odot}/yr)$ & $8.1 \cdot 10^{-8}$ & $1.8 \cdot 10^{-9}$ \\
$L_{nuc,g}/L_{surface,g}$ & \multicolumn{2}{c}{$\sim 0.98$} \\
$X_{c,g}$ & 0.70 & 0.48 \\
\hline
\end{tabular}
\end{table*}

The different behavior of the two groups of systems can be understood in terms of the accretion timescale $M_g / \dot{M}$ versus the thermal timescale $t_{KH}$ ($\sim M_g^2/RL$). This ratio is always larger than one for the mass ratios between one and two (at maximum mass loss rate it reaches $\sim 6$). However, for large initial mass ratios, the ratio of the timescales becomes as small as 0.01 at maximum $\dot{M}$, while it is $\sim 100$ when $M_g = M_{observed}$.

In the slow phase of mass transfer, all calculated series have gainers that are close to thermal equilibrium. For the large initial mass ratios, the ratio $L_{nuc} / L_{surface} \sim 0.98$ when the mass of the gainer matches the observed one (compared to a value of 0.05 during maximum mass loss). For the three systems with mass ratios between one and two we obtain average values of 0.91 (slow phase) and 0.16 (maximum mass loss rate). When accretion starts, since the thermal timescale of the donor is much smaller than for the gainer, the core is compressed almost adiabatically. As a result the temperature and density rise and generate an energy surplus in the interiors. This results in an increase of $L$ and $R$. The difference between the stars is in the ability of their external layers to absorb this flux of energy. Similar results were found by \cite{webbink76} and \cite{webbink77}. In more massive stars, thermal timescales are similar and the gainer evolves more or less along the TAMS, while in lower mass stars it deflects from the ZAMS.

Hence, to obtain an accretion model near the observed HRD position of the gainer, an initial mass ratio of $\sim 3.4$ is required. However, as the figure shows, during the fast mass transfer the gainer moves almost vertically in the HRD, obtaining a radius that largely exceeds the Roche radius. The conservative evolution beyond the point where $R_g = R_{crit}$ is therefore formal. Thus, a non-conservative (liberal) approach is necessary.

The other two systems give similar conclusions. To reach the observed luminosity of the observed gainer, one has to start from a large mass ratio. The position of RZ~Cas is reached with $q_i = 2.07$ ($M_{gi} = 0.705 M_{\odot}$), but then the model mass is only $1.89 M_{\odot}$. The observed mass is reached for $q_i = 2.94$ ($M_{gi} = 0.705 M_{\odot}$), but with $\mathrm{log} \ T_{eff} = 3.98$ and $\mathrm{log} \ L / L_{\odot} = 1.27$. Here also, a large radius excess over the Roche radius occurs during the first and fast phase of mass transfer. Similarly, for S~Equ a value of $q_i \sim 3.3 - 3.4$ is minimally needed, with even greater radius excess. For S~Equ an additional problem occurs with conservative evolution. The large mass ratio results in large values for $M_{di}$, with radii that are already matching the Roche radii at the beginning of the main sequence evolution.

The observed underluminosity remains a problem. The three systems with mass ratio between one and two were stopped at mass $M_g = 1.80 M_{\odot}$ to $1.86 M_{\odot}$. They turn out to be slightly underluminous. Taking for example the middle system, $M_{gi} = 1.32 M_{\odot}$, we obtain at $M_g = 1.84 M_{\odot}$ that $\mathrm{log} \ L_g/L_{\odot} = 1.09$ ($X_c = 0.54$) whereas a $1.84 M_{\odot}$ single star has $\mathrm{log} \ L/L_{\odot} = 1.12$ at the same $X_c$. However, in the large mass ratio series we find $\mathrm{log} \ L/L_{\odot} \sim 1.6 - 1.7$ at $M_g = 2.53 M_{\odot}$ (equal to that of a single star of $2.53 M_{\odot}$), whereas KO~Aql has $\mathrm{log} \ L/L_{\odot} = 1.4$.

\section{Conclusion}

We have found that in order to reproduce the gainers of RZ~Cas, KO~Aql and S~Equ through conservative binary evolution, one would have to start from large initial mass ratios (typically $> 3$). The corresponding high mass loss rates in the beginning of the RLOF result in radii larger than the Roche radii. Hence, a non-conservative approach is needed, in which processes are considered that enhance the period, without losing too much mass (as this would result in a higher initial mass ratio and problems in reaching the final mass and luminosity in the accretion process). We have planned to undertake such a study in the coming months, exploring also the reasons for the observed underluminosity of the gainers.

\begin{acknowledgements}
      LRY acknowledges warm hospitality and support from VUB and support by RFBR grant 07-02-00454 and Russian Academy of Sciences Basic Research Program ``Origin and Evolution of Stars and Galaxies''. We thank the anonymous referee for valuable comments.
\end{acknowledgements}

\end{document}